\documentclass[epj]{svjour}
\usepackage{graphics}
\usepackage{amsfonts,amsmath,amssymb,bm}
\usepackage{dcolumn}
\usepackage{epsfig}
\begin{document}

\title{Scale free effects in world currency exchange network}

\author{A. Z. G\'orski\inst{1} \and S. Dro\.zd\.z\inst{1,2}
\and J. Kwapie\'n\inst{1}}
\institute{Institute of Nuclear Physics, Polish Academy of Science, PL--31-342
Krak\'ow, Poland
\and Institute of Physics, University of Rzesz\'ow, PL--35-310 Rzesz\'ow,
Poland}
\date{Received: date / Revised version: date}
\abstract{A large collection of daily time series for 60 world currencies' 
exchange rates
is considered. The correlation matrices are calculated and the corresponding 
Minimal
Spanning Tree (MST) graphs are constructed for each of those currencies used 
as reference
for the remaining ones. It is shown that multiplicity of the MST graphs' 
nodes to a good
approximation develops a power like, scale free distribution with the scaling 
exponent
similar as for several other complex systems studied so far. Furthermore, 
quantitative
arguments in favor of the hierarchical organization of the world currency 
exchange
network are provided by relating the structure of the above MST graphs and 
their scaling
exponents to those that are derived from an exactly solvable hierarchical 
network model.
A special status of the USD during the period considered can be attributed to 
some
departures of the MST features, when this currency (or some other tied to it) 
is used as
reference, from characteristics typical to such a hierarchical clustering of 
nodes
towards those that correspond to the random graphs. Even though in general 
the basic
structure of the MST is robust with respect to changing the reference 
currency some trace
of a systematic transition from somewhat dispersed -- like the USD case -- 
towards more
compact MST topology can be observed when correlations increase.
\PACS{
{89.65.Gh}{Economics; econophysics, financial markets, business and management} \and
{89.75.Fb}{Structures and organization in complex systems} \and
{05.45.Tp}{Time series analysis}
}
}

\maketitle

The world currency exchange market (foreign exchange -- FOREX, FX) is the 
world largest
financial market and it constitutes an extremely complex network. The FX 
daily takeover
volume is of the order of $10^{12}$ USD. Any other financial market can 
hardly approach
such volume. Also, this market has direct influence on all other markets 
because any
price is expressed in terms of a currency. The large volume makes it 
virtually impossible
to control from outside and there is no friction (transactions are basically 
commission
free). Due to time differences FX transactions are performed 24 hours a day, 
5.5 day a
week with maximum volume between 1 and 4 p.m. GMT, when both American and 
European
markets are open. Hence, the FX time series relations represent an exceptionally 
complex network
indeed, and they are therefore especially worth of detailed analysis. The FX 
market can
be viewed as a complex network of mutually interacting nodes, each node being 
an exchange
rate of two currencies. In principle, all nodes are interconnected with 
complex nonlinear
interactions. Any currency can be expressed in terms of particular one that 
is called the
{\em base currency}. In spite of its importance, much less attention has been 
paid in
literature to the FX cross-correlation analysis, than to such analysis of 
stock
markets\cite{Laloux1999,Stock1,Stock2}. Therefore in the following the 
currency network
will be analyzed. Our motivation to investigate correlations of FX time 
series is
twofold: theoretical and practical. More detailed correlation analysis can 
give insight
into the structure of links between various currencies and, in particular, it 
potentially
may provide quantitative arguments in favor of an often postulated 
hierarchical
organization of world currency exchange market. Knowledge of correlations is 
also
essential for the portfolio management.

Usually, for a financial time series of an i$th$ asset ($i=1, \ldots, n$) at 
time $t$,
$x_i(t)=x_i$, one defines its return over time period $\tau$ as $G_i(t; \tau) 
= \ln
x_i(t+\tau) - \ln x_i(t)$. For FX series instead of a value $x_i(t)$ one has 
$x_A^B(t)$,
an exchange rate, {\em i.e.} a value of currency $A$ expressed in terms of a 
base
currency $B$. Hence, the returns can be denoted as $G_A^B(t)$ and they are 
clearly
antisymmetric $G_A^B(t; \tau) = - G_B^A(t; \tau)$. FX returns, due to the 
lack of
commission and high liquidity satisfy the {\it triangle rule}: $G_A^B(t; 
\tau) + G_B^C(t;
\tau) + G_C^A(t;\tau) = 0$ , already for relatively small values of
$\tau$\cite{Aiba2003}. As a result, for a set of $n$ currencies we have $N=n-1
$
independent values and the same number of nodes with a given base currency.

In the following we analyze time series of daily data for 60 currencies, 
including gold,
silver and platinum\cite{DataSource}. The data taken covers the time period 
Dec 1998--May
2005. In order to automatically get rid of possible misprints in the original 
data the
daily jumps greater than $5\sigma$ (less than 0.3\% of data points) were 
removed. Also,
the gaps related to non-trading days were synchronized. For each exchange 
rate we thus
obtain a time series of 1657 data points. The currencies are denoted 
according to ISO
4217 standard, and they can be formally divided into four groups, according 
to their
liquidity. The major currencies, that we call the $A^\star$ group, include USD, 
EUR, JPY, GBP,
CHF, CAD, AUD, NZD, SEK, NOK, DKK (11 currencies). All other liquid currencies 
belong to the group $A$ (CYP, CZK, HKD, HUF, IDR, ILS, ISK, KRW, MXN, MYR, PHP, 
PLN, SGD, SKK, THB, TRY, TWD, XAG, XAU, XPT, ZAR, 21 currencies). 
Less liquid currencies (group $B$) include: ARS, BGN, BRL, CLP, KWD, RON, RUB, SAR, TTD 
(9 currencies). Finally, the
non-tradable currencies (group $C$) taken into account are: AED, COP, DZD, 
EGP, FJD, GHC,
HNL, INR, JMD, JOD, LBP, LKR, MAD, PEN, PKR, SDD, TND, VEB, ZMK (19 
currencies). In the
latter group the exchange rates are usually fixed daily by national central 
banks.
Dividing currencies into such four groups of different liquidity is common 
among finance
practitioners. This therefore opens an additional interesting issue to be 
verified if the
different dynamics that stays behind such a division according to the 
liquidity (implying
the fixing method) is also reflected in correlations of daily exchange rates.

For a given choice of the base currency $X$ the (symmetric) correlation 
matrix (CM) can
be computed in terms of the normalized returns, $g_A^X(t)$. To this end one 
takes $N$ time
series $\{ g^X_A(t_0), g^X_A(t_0+\tau), \ldots, g^X_A(t_0+(T-1)\tau)\}$ of 
length $T$.
These series can form an $N\times T$ rectangular matrix ${\bm M}^X$, and the 
CM can be
written in the matrix notation as
\begin{equation}
\label{CM2}
{\bm C^X} \equiv [C^X]_{AB} = \frac{1}{T} \
{\bm M}^X {\widetilde{\bm M}^X} \ ,
\end{equation}
where tilde stands for the matrix transposition. By construction the trace of 
the CM is
equal to the number of time series $\text{Tr} \ {\bm C^X} = N$. When some of 
the time
series become strongly dependent, the related number of eigenvalues approach 
value zero
({\em zero modes}). This effect, typical for FX data, does not occur for the 
stock market
time series.

Sample spectra of CMs for the same time series as the ones considered here, 
using several
different base currencies $X$, can be found in \cite{EPJ2007}. In all the 
cases one finds
a maximal eigenvalue ($\lambda_{max}$, in our case $max=N=59$) that is well 
separated
from all the remaining eigenvalues and is placed within the range $0.2 N \div 
0.9 N$,
with average around $0.5 N$. With a few exceptions, there are also clearly 
isolated
second maximal eigenvalues ($\lambda_{max-1}$), considerably smaller than
$\lambda_{max}$, but also well separated from the bulk of other eigenvalues. 
At the same
time, the magnitude of the largest eigenvalue significantly depends on which 
currency is
used as a base $X$. The largest maximal eigenvalues correspond to those base 
currencies
that either have a very strong drift, independent of the behavior of other 
currencies
(like GHC) or whose fluctuations are to a large extent independent of the 
global FX
behavior. The smallest values of $\lambda_{max}$ appear when the USD or other 
currencies
that are strongly tied to it are used as $X$. This reflects the fact that 
changes in
value of the USD - due to its global world significance - cause a rich 
diversity of
reactions of all other currencies. In other words, their dynamics viewed from 
the USD
perspective looks less cooperative as compared to the cases when the base 
currency is not
that influential. Thus, effectively eliminating the USD by using it as the 
base currency
sizeably diminishes correlations among all other currencies. Another 
interesting
manifestation of this effect will be seen in the Minimal Spanning Tree (MST) 
graphs
picture. It should be mentioned that analogous characteristics for 
eigenvalues have been
identified in the smaller subsectors (like tradable and non-tradable 
separately) of the
analyzed currencies.

The structure of eigenspectrum is one way to get insight into large amount of 
data as
represented by the correlation matrices. A complementary - efficient and 
conclusive - way
to visualize such ensembles of data is in terms of the MST graphs. In the 
following we
use the MST graphs approach in order to further explore properties of the 
currency
exchange network. The MST was introduced in graph theory quite long ago
\cite{Boruvka1926,Kruskal1956}. Later it was rediscovered several times
\cite{Papadimitru,West96}. To analyze the stock market correlations it was 
applied by
Mantegna \cite{MantegnaMST} and later on by several authors
\cite{Kertesz2003,Bonanno2003,Onella2003,Bonanno2004}. Recently MST graphs 
were used to
study FX correlations\cite{McDonald2005,Gorski2006,Naylor2007}. Here we 
present the most
exhaustive results concerning the FX data, for 60 world currencies.

To construct the MST graph we choose the following measure
\begin{equation}
\label{MSTmetric}
d^X(A,B) = \sqrt{ (1 - C^X_{AB})/2 } \ , \quad d^X(A,B) \in [0, 1] \ ,
\end{equation}
that satisfies the standard axioms of metric. The distance $d$ between two 
time series is
smaller if their correlation coefficient is closer to unity. 
One node (vertex) in the graph corresponds to each time series. 
We shall connect the two nodes, $A$ and $B$, with a line (edge, ``leg''), 
if $d(A,B)$ is the smallest. In the next step we
look for another two closest nodes and again we connect them with a line. 
This procedure
is repeated until we obtain a connected tree graph. 

We call {\em multiplicity} ({\em degree}) of that node the number 
of legs attached to it. 
Clearly, multiplicity of a node is an integer number, $K$. 
In our case we have $N = 59$ nodes for a given base
currency. As examples, Fig.~1 shows the MST graphs for our set of data for 
three major
currencies, USD, GBP and JPY, taken as the base currency.
\begin{figure}
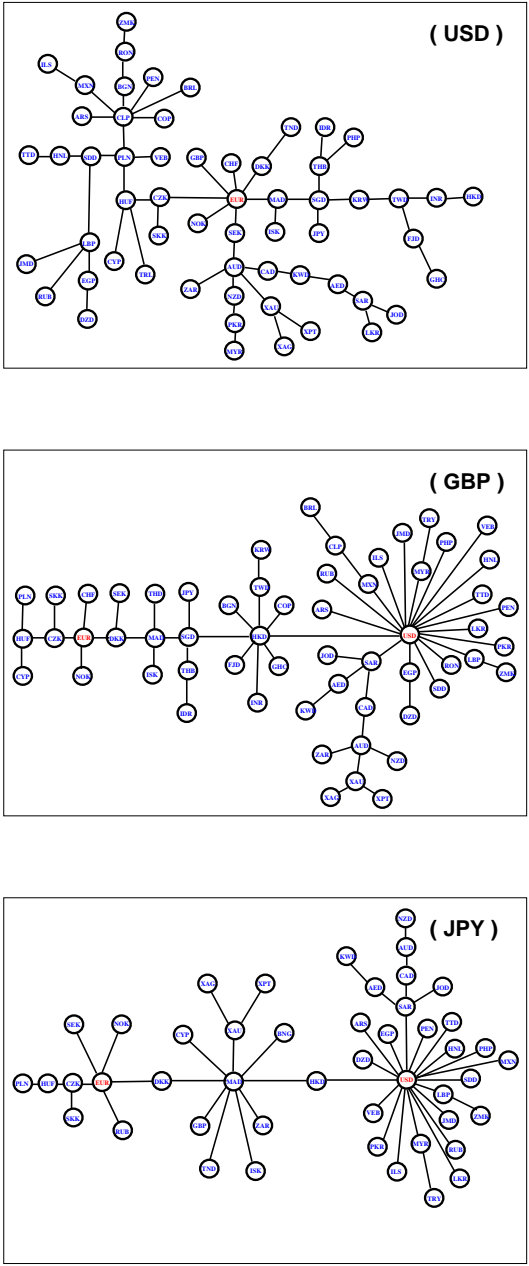

\epsfxsize 7.0cm
\epsffile{usd.eps}

\vspace{1.0cm}
\epsfxsize 7.0cm
\epsffile{gbp.eps}

\vspace{1.0cm}
\epsfxsize 7.0cm
\epsffile{jpy.eps}
\caption{MST graphs for all the currencies taking USD, GBP and JPY as the base
currency, respectively.}
\end{figure}
For a node $A$ its multiplicity is denoted by $K_A$. Integer $N'(K)$ is the number of
nodes with exactly $K$ legs. Because the total number of nodes, $N$, is 
relatively small, we introduce the integrated quantity
\begin{equation}
\label{integratedN} F(K) = N(K)/N = \frac{1}{N}\sum^{K_{max}}_{I=K} N'(I) \ ,
\end{equation}
where $K_{max}$ denotes the maximal number of legs in the MST graph. $N(K)$ 
is the number
of nodes with $K$ or more legs. Counting legs in all nodes of our MST graph 
we can
construct discrete function $N(K)$ and its normalized version, $F(K)$.

\begin{table}
\begin{center}
\caption{\label{tab:table1}
Characteristics of the power fits for the MST node's
multiplicity of all the currencies using the base currencies from the group
$A^*$.}
\begin{tabular}{|lcccc|}%
\hline %
\hline %
 \ base currency & $\alpha$ & $\Delta\alpha$ & $\Delta\alpha/\alpha$
& $\lambda_{max}$\\
\hline%
\hline%
 \ \ \ AUD & 1.43 & 0.08 & 5.4\% & 32.1 \\
\ \ \ CAD & 1.39 & 0.11 & 7.7\% & 23.2 \\
\ \ \ CHF & 1.34 & 0.08 & 6.0\% & 29.9 \\
\ \ \ DKK & 1.33 & 0.08 & 5.9\% & 27.4 \\
\ \ \ EUR & 1.30 & 0.08 & 5.9\% & 27.6 \\
\ \ \ GBP & 1.44 & 0.11 & 7.6\% & 24.1 \\
\ \ \ JPY & 1.36 & 0.08 & 6.0\% & 32.6 \\
\ \ \ NOK & 1.41 & 0.08 & 5.5\% & 28.4 \\
\ \ \ NZD & 1.43 & 0.07 & 5.0\% & 34.4 \\
\ \ \ SEK & 1.41 & 0.07 & 4.7\% & 29.2 \\
\ \ \ USD & 1.92 & 0.18 & 9.6\% & 11.8 \\%
\hline
\ \ \ average & 1.43 & 0.09 & 6.3\% & 27.3 \\
\hline
\hline
\end{tabular}
\end{center}
\end{table}
In Fig.~2 we show the $F(K)$ plots for the MST node's multiplicity for 
different base
currencies from each of the four groups, $A, A^\star, B$ and $C$. To make the 
plots
readable we present the distributions for five typical basic currencies from 
each group.
Already at first glimpse, basically all plots indicate scaling and their 
similarity
extends even to a quantitative level. One subtle difference can be visible 
when the USD
(as well as the other currencies linked to it) is used as reference. Here, in 
the log-log scale one observes a somewhat faster decrease than linear. 
Nevertheless, fitting it with
the straight line results thus in a steeper slope and the worse quality of 
such a fit.
The relevant detailed numerical parameters, including the power exponent 
$\alpha$, its standard error ($\Delta\alpha$), and values of the maximal 
CM eigenvalues, for all the 60
currencies used as the base for the remaining ones are listed in Tables~1-4. 
The exponents $\alpha$ and their standard errors $\Delta\alpha$ were determined 
with the least square fit of corresponding power functions in the original linear 
scale. Because the total number of data points equals 59 and the number of occupied bins is 
typically around 10 more sophisticated statistical estimates do not seem adequate in this case.

\begin{table}
\begin{center}
\caption{\label{tab:table2}
Characteristics of the power fits for the MST node's
multiplicity of all the currencies using the base currencies from the group
$A$.}
\begin{tabular}{|lcccc|}%
\hline %
\hline %
\ base currency & $\alpha$ & $\Delta\alpha$ &
$\Delta\alpha/\alpha$ & $\lambda_{max}$\\
\hline%
\hline%
 \ \ \ CYP & 1.42 & 0.07 & 4.6\% & 36.6 \\
\ \ \ CZK & 1.33 & 0.08 & 5.8\% & 31.7 \\
\ \ \ HKD & 2.31 & 0.32 & 14.0\% & 11.7 \\
\ \ \ HUF & 1.38 & 0.08 & 5.8\% & 31.8 \\
\ \ \ IDR & 1.48 & 0.09 & 5.9\% & 43.7 \\
\ \ \ ILS & 1.56 & 0.12 & 7.9\% & 24.2 \\
\ \ \ ISK & 1.47 & 0.09 & 5.9\% & 27.6 \\
\ \ \ KRW & 1.59 & 0.11 & 6.7\% & 27.8 \\
\ \ \ MXN & 1.51 & 0.07 & 4.3\% & 29.8 \\
\ \ \ MYR & 2.25 & 0.31 & 14.0\% & 11.9 \\
\ \ \ PHP & 1.62 & 0.11 & 6.7\% & 24.5 \\
\ \ \ PLN & 1.49 & 0.08 & 5.7\% & 32.0 \\
\ \ \ SGD & 1.67 & 0.10 & 5.9\% & 13.8 \\
\ \ \ SKK & 1.41 & 0.06 & 4.4\% & 31.7 \\
\ \ \ THB & 1.50 & 0.11 & 7.3\% & 20.5 \\
\ \ \ TRY & 1.55 & 0.08 & 5.1\% & 43.3 \\
\ \ \ TWD & 1.61 & 0.12 & 7.7\% & 15.2 \\
\ \ \ XAG & 1.48 & 0.11 & 7.7\% & 46.7 \\
\ \ \ XAU & 1.38 & 0.08 & 5.5\% & 37.2 \\
\ \ \ XPT & 1.40 & 0.09 & 6.3\% & 47.1 \\
\ \ \ ZAR & 1.50 & 0.08 & 5.7\% & 41.8 \\%
\hline
\ \ \ average & 1.57 & 0.11 & 6.8\% & 30.0 \\
\hline
\hline
\end{tabular}
\end{center}
\end{table}
\begin{table}
\begin{center}
\caption{\label{tab:table3} Characteristics of the power fits for the
 MST node's multiplicity of all the currencies using the basic currencies
from
the group $B$.}
\begin{tabular}{|lcccc|}%
\hline %
\hline %
\ base currency & $\alpha$ & $\Delta\alpha$ &
$\Delta\alpha/\alpha$ & $\lambda_{max}$\\
\hline%
\hline%
 \ \ \ AED & 1.92 & 0.13 & 6.5\% & 17.5 \\
\ \ \ ARS & 1.45 & 0.06 & 4.4\% & 40.2 \\
\ \ \ BGN & 1.44 & 0.09 & 5.3\% & 32.3 \\
\ \ \ BRL & 1.60 & 0.11 & 6.7\% & 43.5 \\
\ \ \ CLP & 1.48 & 0.09 & 6.0\% & 31.0 \\
\ \ \ KWD & 1.85 & 0.11 & 6.1\% & 17.9 \\
\ \ \ RON & 1.50 & 0.08 & 5.0\% & 36.7 \\
\ \ \ RUB & 1.48 & 0.09 & 5.8\% & 44.5 \\
\ \ \ SAR & 1.65 & 0.09 & 5.3\% & 16.9 \\
\ \ \ TTD & 1.97 & 0.14 & 7.1\% & 15.6 \\
\hline %
\ \ \ average & 1.63 & 0.09 & 5.8\% & 29.6 \\
\hline %
\hline %
\end{tabular}
\end{center}
\end{table}
\begin{table}
\begin{center}
\caption{\label{tab:table4} Characteristics of the power fits for the MST
node's multiplicity of all the currencies using the base currencies from
the group $C$.}
\begin{tabular}{|lcccc|}%
\hline %
\hline %
\ base currency & $\alpha$ & $\Delta\alpha$ &
$\Delta\alpha/\alpha$ & $\lambda_{max}$\\
\hline%
\hline%
 \ \ \ COP & 1.46 & 0.07 & 7.1\% & 30.8 \\
\ \ \ DZD & 1.44 & 0.08 & 6.3\% & 51.3 \\
\ \ \ EGP & 1.50 & 0.32 & 4.7\% & 39.5 \\
\ \ \ FJD & 1.48 & 0.08 & 6.0\% & 31.0 \\
\ \ \ GHC & 1.54 & 0.09 & 4.9\% & 52.2 \\
\ \ \ HNL & 1.81 & 0.12 & 5.0\% & 14.0 \\
\ \ \ INR & 1.92 & 0.09 & 5.5\% & 14.3 \\
\ \ \ JMD & 1.66 & 0.09 & 9.0\% & 35.0 \\
\ \ \ JOD & 1.80 & 0.07 & 5.5\% & 20.6 \\
\ \ \ LBP & 1.31 & 0.31 & 9.0\% & 24.6 \\
\ \ \ LKR & 1.73 & 0.11 & 7.3\% & 18.5 \\
\ \ \ MAD & 1.37 & 0.08 & 11.0\% & 21.1 \\
\ \ \ PEN & 1.78 & 0.10 & 7.5\% & 17.4 \\
\ \ \ PKR & 1.70 & 0.06 & 4.7\% & 25.1 \\
\ \ \ SDD & 1.67 & 0.11 & 5.8\% & 25.8 \\
\ \ \ TND & 1.39 & 0.08 & 8.0\% & 21.6 \\
\ \ \ VEB & 1.44 & 0.12 & 5.8\% & 37.6 \\
\ \ \ ZMK & 1.31 & 0.11 & 7.7\% & 49.3 \\
\hline%
\ \ \ average & 1.57 & 0.11 & 6.7\% & 29.4 \\
\hline%
\hline%
\end{tabular}
\end{center}
\end{table}
\begin{figure*}
\begin{center}
\includegraphics[width=16.0cm,angle=0]{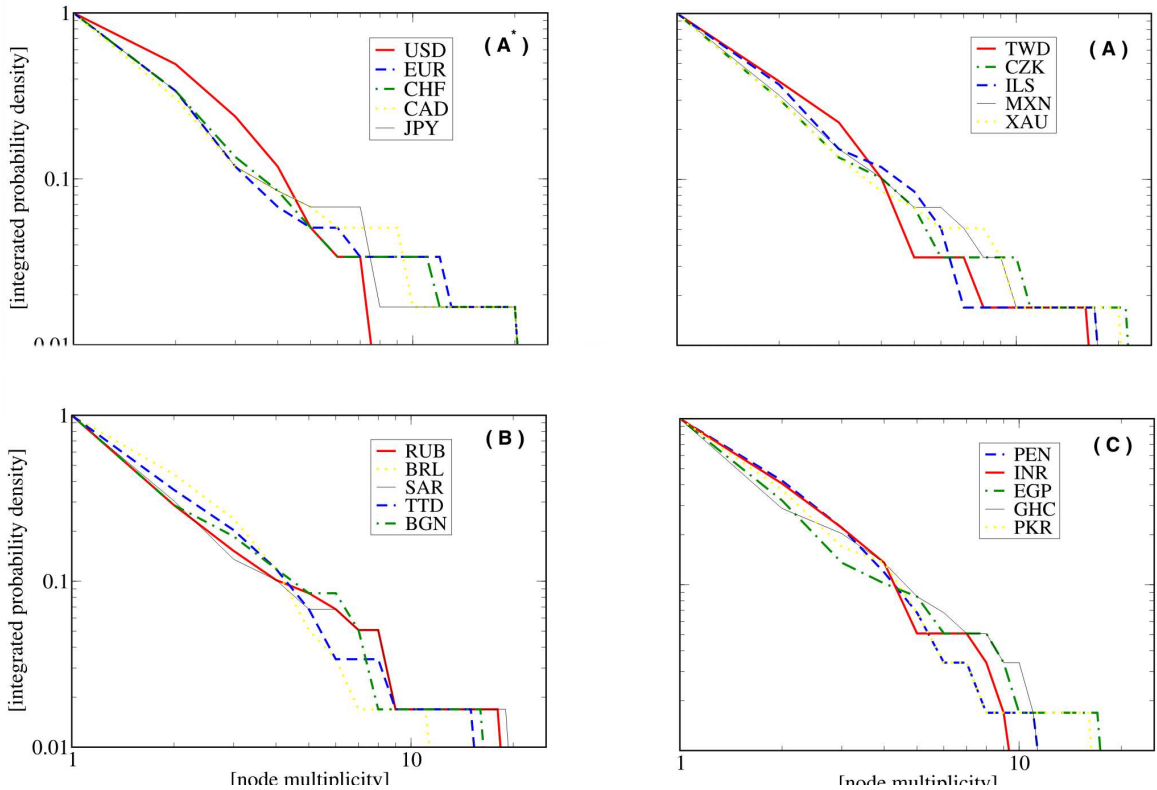}
\caption{Log-log plots of the function $F(K)$ for five sample base
currencies taken from each group:
$A^*$, $A$, $B$ and $C$, respectively.}
\end{center}
\label{fig:fig4}
\end{figure*}

It can be concluded that the scale free behavior and thus the power like fit 
are fairly good for majority of the base currencies. The worst scaling (if at all),
$\Delta\alpha/\alpha \ge 9$\%, was obtained for the USD and currencies tied 
to it (HKD, MYR, LBP). 
A technical reason for the latter effect is, that taking USD as the
base currency one eliminates its node from the MST graph. Hence, the largest 
cluster
(i.e. large multiplicity node) disappears, the tail of the distribution is 
getting
thinner and the MST graph is too small to be able to reproduce fat tails 
(i.e. power like
distribution). On a more fundamental level, this characteristic feature 
associated with
the USD -- as seen from the MST perspective -- may reflect its special status 
in the
world economy during the period considered here. 
It is probably also worth to notice that the quality of scaling goes somewhat 
in parallel with the magnitude of the largest eigenvalue (last column in the Tables) 
of the original correlation matrix.
The disputable -- as far as the scaling is concerned -- cases are typically 
associated
with the relatively small magnitudes of $\lambda_{max}$ and thus with the 
lower degree of
collectivity.

For the prevailing majority of the base currencies the statistical 
significance of the
scaling effects is however quite convincing. Even more, the corresponding 
scaling
exponents do not differ much among the four groups of currencies considered. 
The scale
free behavior of MST graphs is in agreement with the successful usage of 
hierarchical
structure methods in finance \cite{Naylor2007} and the fat tails are 
signature of
currency clustering in MST graphs. In more quantitative terms, the power like 
scaling of
nodes' degree has been shown in hierarchical networks \cite
{Ravasz2003,Noh2003} and the
corresponding scaling exponents derived analytically. In the context of our 
present study
it is very interesting and encouraging to find out that the average value we 
have
obtained for the tail index is $\langle \alpha \rangle = 1.55$ while the 
hierarchical
model discussed in \cite{Ravasz2003,Noh2003} implies very close value (for the 
cumulative
distribution)
\begin{equation}
\label{TheorIndex} \alpha_{hier} = \ln M / \ln (M-1) = 1.61 \ .
\end{equation}
The replication factor $M$ to be used in this formula in our case can be 
calculated from
the average node degree: $M = \langle K \rangle + 1 = 2 (N-1)/N + 1 = 2.966$. 
The scaling
exponent (\ref{TheorIndex}) thus differs from our average value 
$\langle \alpha \rangle$
by less than $4$\%. This fact provides a significant quantitative argument in 
favor of
the hierarchical organization of the world currency exchange market, an 
effect which
already can visually be inferred from the panels corresponding to GBP and JPY 
in Fig.~1.

The panel corresponding to the USD looks less explicit in this respect. 
Indeed the quality of the corresponding scaling is here not so convincing.
Nevertheless, approximation with the power like fit gives the exponent 
$\alpha$ which is significantly larger (Table~1) than in the above 
hierarchical model. 
In fact, the node's multiplicity distribution corresponding to
this case and seen in Fig~1, develops some departure towards a Gaussian 
distribution and thus indicates some admixture of randomness and a more 
"democratic" multiplicity distribution as compared to a pure hierarchical situation. 
This effect also is consistent -- as expressed by the corresponding $\lambda_{max}$ 
in Table~1 -- with the weakest synchronization of the currency exchange rates expressed 
in the USD. 
Interestingly, a correspondence of this kind seems to be leaving imprints for 
the other currencies when they are used as reference for all remaining. 
As it can be seen from Tables~1-4 the smaller values of $\lambda_{max}$ 
(and thus weaker correlations) are quite systematically
associated with the scaling exponents $\alpha$ which are larger than their 
average value
$\langle \alpha \rangle = 1.55$ while the larger values of $\lambda_{max}$ 
(stronger
correlations) are typically associated with $\alpha$ smaller than this 
average. For a
better visualization the quality of such a tendency is demonstrated in Fig.~3 
by the
scatter plot of $\alpha$ versus the corresponding $\lambda_{max}$ for all the 
60
currencies. An overall decrease of $\alpha$ with increasing $\lambda_{max}$ 
is quite
evident from this Figure. It is faster in the region of smaller 
$\lambda_{max}$ values
and then tends to saturation. The best simple representation for this 
dependence (solid
line in Fig.~3) is in terms of a power function 
$(\alpha \sim (\lambda_{max}-\lambda_{max}^{RM})^{-\beta})$
with $\beta \approx 0.21$. 
As $\lambda_{max}^{RM}$ the upper bound of the Wishart ensamble of random matrices 
is used. Thus 
$\lambda_{max}^{RM} = 1 + 1/Q + 2/\sqrt{Q}$ \cite{MarchenkoPastur,SenguptaMitra}, 
where $Q = T/N$ which in our case gives $\lambda_{max}^{RM} = 1.413$.

\begin{figure}
\begin{center}
\epsfxsize 8.5cm
\epsffile{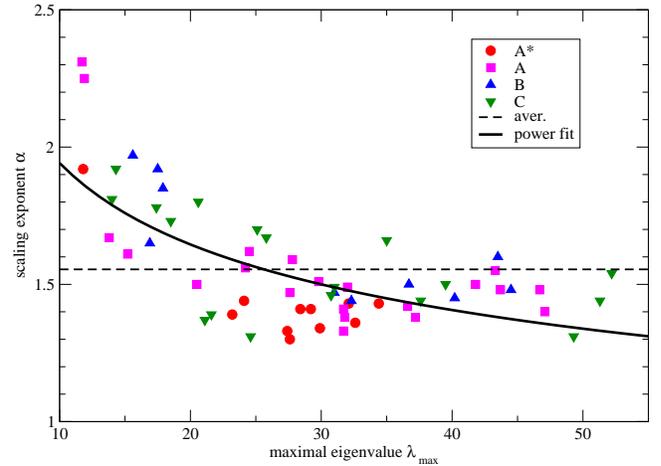}
\caption{Scatter plot of $\alpha$ versus $\lambda_{max}$ for all the 60 
currencies. Four different symbols correspond to four different groups
of the currencies as described in the inset. The solid line represents
the best fit in terms of a power function and the dashed horizontal line
indicates the average $\langle \alpha \rangle$. }
\end{center}
\label{fig:fig5}
\end{figure}

The above correspondence between $\lambda_{max}$ and $\alpha$ remains in
accord and in fact constitutes an extension of the observation made in
ref.~\cite{Onella2003} in connection with the stock market. 
There it is shown that the MST scaling exponent evaluated during crash 
periods -- whose dynamics is inherently associated with stronger 
correlations~\cite{Drozdz2000} -- is smaller than during normal business
periods. As already mentioned and even more extensively discussed
in~\cite{EPJ2007} the currencies exchange rates expressed in terms 
of a less influential currency are more correlated as compared 
to the situation when the leading ones are used as reference. 
It is also interesting to see that the above coincidences apply 
to all the four groups of currencies listed in Tables~1-4 and displayed
in Fig.~3 (where they are seen to overlap with each other) even though 
they are characterized by sizeably different liquidity and in some
cases (group C) even by non-trade mechanism of setting the exchange rates. 
This implies
that the general characteristics of the corresponding MST's are quite robust 
with respect
to such factors.

The power like behavior with the scaling exponent $\alpha = 1.6$, which is 
very close to
the one estimated in the present study, was also found~\cite{Bonanno2003} for 
the stock market MST graphs. 
Furthermore, similar characteristics have been identified 
in several other natural and artificial complex networks studied so far 
in the literature -- for references see \cite{alphaExmpl1,alphaExmpl2} --
and the world currency network considered here belongs 
to this universality class, 
with the scaling exponent in the range $1 < \alpha < 2$.

\end{document}